\documentclass[journal=jpccck,manuscript=article]{achemso}

\usepackage[version=3]{mhchem}
\usepackage[T1]{fontenc} 

\author{Oleg Rubel}
\affiliation{Department of Materials Science and Engineering, McMaster University, 1280 Main Street West, Hamilton, Ontario L8S 4L8, Canada}
\email{rubelo@mcmaster.ca}
\author{Thuy Nguyen Thanh Tran}
\affiliation{Department of Chemical and Materials Engineering, University of Alberta, 9211 116 Street NW, Edmonton, Alberta, Canada T6G 1H9}
\alsoaffiliation{Salient Energy Inc., 21 McCurdy Ave, Dartmouth, Nova Scotia, Canada B3B 1C4}
\author{Storm Gourley}
\affiliation{Department of Chemical Engineering, McMaster University, 1280 Main Street West, Hamilton, Ontario, Canada L8S 4L8}
\author{Sriram~Anand}%
\affiliation{Department of Materials Science and Engineering, McMaster University, 1280 Main Street West, Hamilton, Ontario L8S 4L8, Canada}
\alsoaffiliation{Department of Metallurgical and Materials Engineering, National Institute of Technology Tiruchirappalli, Tamil Nadu 620015, India}
\author{Andrew~Van~Bommel}
\affiliation{Salient Energy Inc., 21 McCurdy Ave, Dartmouth, Nova Scotia, Canada B3B 1C4}
\author{Brian D. Adams}
\affiliation{Salient Energy Inc., 21 McCurdy Ave, Dartmouth, Nova Scotia, Canada B3B 1C4}
\email{brian@salientenergy.ca}
\author{Douglas G. Ivey}
\affiliation{Department of Chemical and Materials Engineering, University of Alberta, 9211 116 Street NW, Edmonton, Alberta, Canada T6G 1H9}
\email{divey@ualberta.ca}
\author{Drew~Higgins}
\affiliation{Department of Chemical Engineering, McMaster University, 1280 Main Street West, Hamilton, Ontario, Canada L8S 4L8}
\email{higgid2@mcmaster.ca}

\title[Electrochemical Stability of \ce{ZnMn2O4}]
  {Electrochemical Stability of \ce{ZnMn2O4}: Understanding Zn-ion Rechargeable Battery Capacity and Degradation}

\begin{document}

\begin{tocentry}

\includegraphics{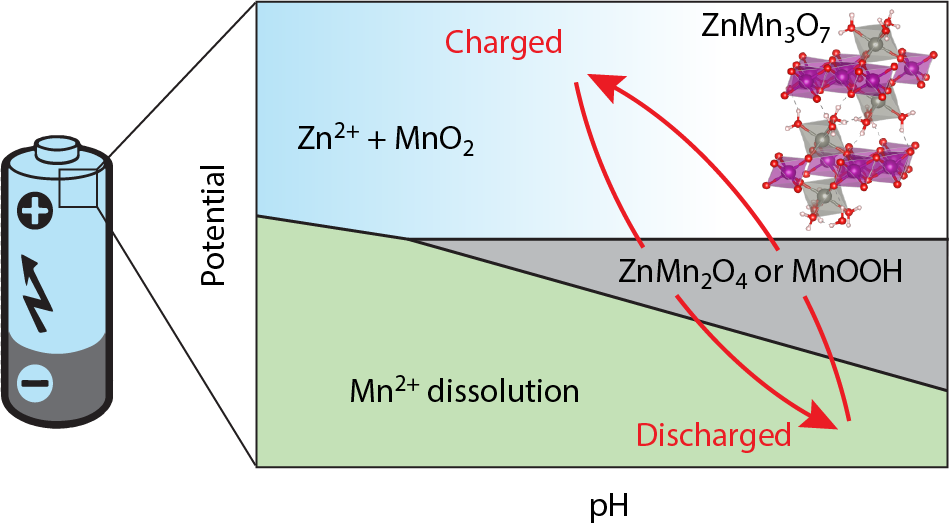}

\end{tocentry}

\begin{abstract}
We present a refined \ce{Mn-Zn-H2O} Pourbaix diagram with the emphasis on parameters relevant for the \ce{Zn/MnO2} rechargeable cells. It maps out boundaries of electrochemical stability for \ce{MnO2}, \ce{ZnMn2O4}, \ce{ZnMn3O7}, and \ce{MnOOH}. The diagram helps to rationalize experimental observation on processes and phases occurring during charge/discharge, including the position of charge/discharge redox peaks and capacity fade observed in rechargeable aqueous Zn-ion batteries for stationary storage. The proposed Pourbaix diagram is validated by observing the pH-dependent transformation of electrolytic manganese dioxide to hetaerolite and chalcophanite during discharge and charge, respectively. Our results can guide the selection of operating conditions (the potential range and pH) for existing aqueous \ce{Zn/MnO2} rechargeable cells to maximise their longevity. In addition, the relation between electrochemical stability boundaries and operating conditions can be used as an additional design criterion in exploration of future cathode materials for aqueous rechargeable batteries.
\end{abstract}

\section{Introduction}\label{sec:Introduction}

Eco-friendly rechargeable Zn-ion batteries with aqueous electrolyte have emerged as a cheaper and safer alternative to Li-ion batteries for certain applications such as grid scale energy storage \cite{Blanc_J_4_2020}. In the first realization of an aqueous  rechargeable Zn-ion battery, \ce{MnO2} was used as the cathode (positive electrode) material \cite{Shoji_JAE_18_1988}. Since then, polymorphs of \ce{MnO2} remain a material of choice for fabrication of practical cells in spite of a variety of alternative positive electrode materials reported to date \cite{Blanc_J_4_2020}. Reversible \ce{Zn^{2+}} intercalation/deintercalation (discharging/charging) in the \ce{MnO2} host framework and formation of \ce{ZnMn2O4} combined with \ce{H+} co-intercalation are recognized as the main energy storage mechanisms in \ce{Zn/MnO2} cells \cite{Xu_ACIE_51_2011,Sun_JACS_139_2017,Zhao_S_15_2019}.

Cycling performance of the \ce{Zn/MnO2} cell is limited by capacity fade, especially at slower cycling rates \cite{Li_CM_31_2019} that are necessary for grid-scale energy storage applications. This capacity fade is attributed to dissolution of the active cathode material into the electrolyte (formation of \ce{Mn^{2+}}(aq)  \cite{Chamoun_ESM_15_2018,Lee_SR_4_2014,Pan_NE_1_2016}) as well as the formation of irreversible phases at the cathode side, such as \ce{Mn(OH)2}, \ce{Mn3O4} or even \ce{ZnMn2O4} \cite{Alfaruqi_EA_276_2018}. Current literature is ambiguous about the electrode  potential where the dissolution takes place. For instance, \citet{Chao_ACIE_58_2019} suggested that \ce{Mn^2+} leaches at high voltages (greater than 1.7~V vs \ce{Zn^0/Zn^{2+}}). On the other hand, \citet{Li_CM_31_2019} attributed capacity fade to reactions that take place at lower voltages (less than 1.26~V vs \ce{Zn^0/Zn^{2+}}). It seems that another solid phase \ce{ZnMn3O7.$n$H2O} (chalcophanite) is also involved in capacity fade \cite{Li_CM_31_2019}. After the initial discharge process, the formation of chalcophanite was observed using X-ray diffraction, and the diffraction peaks associated with this phase became more intense as the cycle number increased, especially at slower cycle rates \cite{Li_CM_31_2019}. Finally, \citet{Li_CM_31_2019} associated the formation of \ce{ZnMn3O7.$n$H2O} and \ce{Mn3O4} below 1.26~V vs \ce{Zn^0/Zn^{2+}} with capacity fade at slow cycling rates. \citet{Tran_SR_11_2021} also reported \ce{ZnMn3O7.3H2O} in \textit{ex situ} studies of Zn/EMD cells (EMD stands for electrolytic manganese dioxide~\cite{Chao_ACIE_58_2019}) and linked the formation of chalcophanite with the electrochemical deposition of dissolved \ce{Mn^{2+}} back on the cathode surface during charging at higher voltages (1.8~V vs \ce{Zn^0/Zn^{2+}}).

While battery operation is governed by non-equilibrium kinetic processes, understanding which phases of the \ce{Mn-Zn-H2O} system are stable within the range of parameters relevant for operation of rechargeable batteries will be important for explaining current stability limitations and designing electrode structures that can provide extended operational lifetimes. In this sense, for aqueous batteries Pourbaix diagrams can be useful for displaying the equilibrium electrochemical stability of a \ce{metal-H2O} system. For example,  \citet{Bischoff_JES_167_2020} created an overlay of \ce{Mn-H2O} and \ce{Zn-H2O} Pourbaix diagrams and measured \textit{in situ} local changes in pH ($3-5$) during the battery cycle within the relevant range of the potential window $0.9-1.9$~V vs \ce{Zn^0/Zn^{2+}}. This diagram allows for several important conclusions: (i) \ce{MnO2} and \ce{Mn2O3} are the only solid phases of the \ce{Mn-H2O} system that are stable in the range specified, (ii) at higher potentials (greater than 1.7~V vs \ce{Zn^0/Zn^{2+}}) the only danger is electrolyte decomposition (via the \ce{O2} evolution reaction) and not cathode dissolution, (iii) during discharge there is a problematic region (below ca.~1~V vs \ce{Zn^0/Zn^{2+}} at pH 5.5) where \ce{Mn^2+}(aq) is more stable than solid \ce{MnO2}. However, the diagram presented by \citet{Bischoff_JES_167_2020} does not consider stability boundaries of two important solids: hetaerolite \ce{ZnMn2O4} and chalcophanite \ce{ZnMn3O7.3H2O}.

\citet{Huang_NL_11_2019} proposed a ternary \ce{Mn-Zn-H2O} Pourbaix diagram, yet it does not have many details in the electrochemical potential region of interest for Zn-ion batteries. Qualitatively, the diagram shows the existence of \ce{ZnMn2O4} and \ce{Zn2Mn3O8}, but does not present clear boundaries for those phases. Also, \ce{ZnMn3O7.3H2O} does not appear on this diagram even though it was extensively discussed in the text of Ref.~\citenum{Huang_NL_11_2019}. It remains unclear what role these Zn-containing phases play in energy storage with respect to the operation of rechargeable Zn-ion batteries with aqueous electrolyte.

To elucidate the electrochemical stability of \ce{ZnMn2O4}, we have provided detailed \ce{Mn-Zn-H2O} Pourbaix analysis focusing on the relevant range of parameters that are encountered in state of the art rechargeable Zn-ion batteries employing a Zn-sulfate electrolyte \cite{Bischoff_JES_167_2020}: namely, pH~$4-6$ and $E_{\ce{Zn^0/Zn^{2+}}}=1.1-1.8$~V. The new diagram outlines the electrochemical potential and pH regions of \ce{ZnMn2O4}, \ce{ZnMn3O7.3H2O}, and \ce{MnOOH} stability. Conclusions drawn from the Pourbaix diagram are corroborated by experimental measurements of the cyclic performance of Zn/EMD cells at variable discharge potentials combined with \textit{ex situ} studies of the cathode material, which provides critical fundamental insight to help understand the stability limitations of conventional electrode structures and guide improved material designs or operational parameters.

\section{\label{sec:Method}Methods}

\subsection{Pourbaix diagram}

Even though construction of Pourbaix diagrams based on electronic structure calculations has become increasingly accurate  \cite{Jain_PRB_84_2011,Persson_PRB_85_2012,Zeng_JPCC_119_2015,Wang_nCM_6_2020}, it is still not fully \textit{ab initio} (mostly hindered by correlation effects of transition metals and finite-temperature thermodynamic properties). Here we build the Pourbaix diagram based on experimental data following the method established by \citet{Pourbaix1974} and use a density functional theory (DFT) only when there are no experimental data. A key ingredient for construction of Pourbaix diagrams is standard chemical potentials $\mu^\circ$ of species, listed in Table~\ref{tab-Chem-pot} at standard conditions per formula unit (f.u.).

\begin{table}[t]
\caption{Chemical potentials of species involved in reactions. Bold font highlights primary values used in calculations in the case of multiple values reported in the literature.}
\label{tab-Chem-pot}
\begin{tabular}{ l c c }
	\hline
    Species & $\mu^\circ$ (eV/f.u.) & Ref. \\
	\hline
    \ce{Zn^{2+}}(aq) & $-1.526$ & \citenum{Pourbaix1974} \\
    \ce{Mn^{2+}}(aq) & $-2.359$ & \citenum{Pourbaix1974} \\
    \ce{MnO4^-}(aq) & $-4.657$ & \citenum{Pourbaix1974} \\
    \ce{H2O} (l) & $-2.458$ & \citenum{Pourbaix1974} \\
    \ce{MnO2} (pyrolusite) & $-4.83$ & \citenum{Jacob_HTMPL_30_2011} \\
    \ce{MnOOH} (manganite) & $\mathbf{-5.81}$ & \citenum{Bratsch_JPCRD_18_1989} (estimated from $E^{\circ}=0.98$~V at pH~0) \\
     & $-5.88$ & \citenum{Huang_NL_11_2019} (Table~S7 in supp. inf.)\\
    \ce{ZnO} & $-3.292$ & \citenum{Maier_JACS_48_1926}\\
    \ce{ZnO} hydr. & $-3.336$ & \citenum{Pourbaix1974}\\
    $\alpha$-\ce{Mn2O3} & $-9.206$ & \citenum{Pourbaix1974} \\
    \ce{ZnMn2O4} (hetaerolite) & $\mathbf{-12.61}$ & \citenum{Hem_GCA_51_1987} \\
     & $-12.72$ & \citenum{Huang_NL_11_2019} (supp. inf.)\\
    \ce{ZnMn3O7} (chalcophanite) hydr. & $-17.83-\delta\mu$ & $\mu^{\circ}(\ce{ZnO}~\text{hydr.})+3\mu^{\circ}(\ce{MnO2})-\delta\mu$ \\
    \hline
\end{tabular}
\end{table}

In our calculations we assume concentrations of aqueous species at 2~M for \ce{Zn^{2+}} and 0.1~M for \ce{Mn^{2+}}. These values are based upon typical concentrations of \ce{ZnSO4} and \ce{MnSO4} used in experimental Zn-ion battery research \cite{Pan_NE_1_2016,Chamoun_ESM_15_2018,Bischoff_JES_167_2020}. The concentration of aqueous species is taken into account when computing their chemical potential (eV), e.g.,
\begin{equation}\label{eq:mu(Mn2+) with concentration}
    \mu(\ce{Mn^{2+}}) = \mu^{\circ}(\ce{Mn^{2+}}) + 0.0592\log_{10} [\ce{Mn^{2+}}],
\end{equation}
where $[\ce{Mn^{2+}}]$ is the aqueous concentration of \ce{Mn^{2+}} ions. Sample calculations are presented in the Appendix. It should be noted that experiments in this paper were performed using an electrolyte with 1~M concentration of \ce{ZnSO4} (see Sec.~Electrode fabrication). This discrepancy has a minor effect on the calculated Pourbaix diagram and does not affect the main conclusions. 

\subsection{Electronic structure calculations}

DFT \cite{Hohenberg_PR_136_1964,Kohn_PR_140_1965} calculations were performed using the Vienna \textit{ab initio} simulation package \cite{Kresse_PRB_54_1996} (VASP) and projector augmented-wave potentials \cite{Kresse_PRB_54_1996,Kresse_PRB_59_1999,Blochl_PRB_50_1994}. The following potentials were used: Mn\_sv, Zn, O\_h, H\_h. A Perdew-Burke-Ernzerhof \cite{Perdew_PRL_77_1996} (PBE) generalized gradient approximation (GGA) for the exchange-correlation functional was chosen in combination with the \citet{Grimme_JCP_132_2010} (D3) correction to capture long-range van der Waals interactions.

A full structural relaxation was conducted for all simulated compounds, which included relaxation of atomic positions (maximum Cartesian force component of 0.05~eV/{\AA}) and stresses (1~kBar as the maximum component of the stress tensor). A cutoff energy of  $E_{\mathrm{cut}}=875$~eV for the plane-wave expansion was used, which corresponds to the maximum value recommended in pseudopotential files further increased by 25\% above (VASP tag $\text{PREC}=\text{High}$). The Brillouin zone was sampled using a \citet{Monkhorst_PRB_13_1976} shifted $k$ mesh with a density of 20 divisions for every 1~{\AA}$^{-1}$ linear dimension in reciprocal space.

Magnetism was included for structures with manganese and the \ce{O2} molecule. In the \ce{O2} molecule the ordering is ferromagnetic. We tested all possible permutations of antiferromagnetic arrangements in addition to the ferromagnetic ordering for structures with manganese and selected the lowest total energy.

\subsection{Electrode fabrication}\label{Methods:Electrode fabrication}

EMD is a commonly utilized electrode material for Zn-ion batteries \cite{Pan_AAMI_11_2019,Tran_SR_11_2021,Poosapati_AAEM_4_2021} and was selected for the current study. EMD shows significant advantages for practical applications such as low synthesis cost and large-scale commercial availability due to the history of being used as a cathode material in commercial alkaline batteries  \cite{Biswal_RA_5_2015}. Other \ce{MnO2} materials, such as \ce{$\alpha$-MnO2} or \ce{$\beta$-MnO2}, could have been chosen, but the effect on the Pourbaix diagram should be minimal (see discussion in the Results section) and does not affect the key findings of this study.

EMD was obtained from Borman Specialty Materials (formerly Tronox). All chemicals were used as received without further purification. A proprietary slurry cast method was used to prepare EMD electrodes \cite{Adams_patent_2019}. The Zn-ion battery cells were made from a frame of acrylic sheets and thin Ti current collectors. The Zn metal electrode was positioned on top of the cell and separated from the underlying \ce{MnO2} electrode by three layers of filter paper (separator) soaked with electrolyte. The contact area between the electrodes and the electrolyte was 4~cm$^2$.

1~M zinc-sulfate aqueous electrolyte was prepared by adding \ce{ZnSO4} to deionized water and stirring vigorously until completely dissolved. 1~M \ce{ZnSO4} solutions with pH 2.5 and pH 4 were used as the electrolyte for the zinc-ion batteries. \ce{H2SO4} was added as needed to adjust the pH of the zinc sulfate electrolyte.

\subsection{Electrochemical testing and characterization of EMD electrodes}

Electrochemical measurements were carried out with a Biologic SP-300 potentiostat. Batteries were first conditioned at a C-rate of C/10 and cycled at C/5. During the charging process, the batteries were held at 1.8~V for an additional 2~h or until the current density dropped to 8~$\mu$A~mg$^{-1}$, whichever came first. After 5 cycles, EMD electrodes were extracted at fully-discharged (0.9~V) and fully-charged states (1.8~V -- 2~h) for characterization.

The morphologies and compositions of the EMD electrodes were characterized using a scanning electron microscope (Tescan Vega3 SEM), coupled with an energy dispersive X-ray (EDX) spectrometer. X-ray diffraction (XRD) analysis was performed using a Rigaku Ultima IV diffractometer with monochromatic Cu~K$\alpha$ X-radiation (wavelength equal to 1.54~{\AA}) at a scan rate of 5 degrees min$^{-1}$. Transmission electron microscopy (TEM) and selected area electron diffraction (SAED) were performed using a JEOL JEM-ARM200CF TEM/STEM operating at an accelerating voltage of 200~kV.

\section{\label{sec:Results and Discussion}Results and Discussion}

\subsection{Electrochemical stability of hetaerolite and \ce{Zn^{2+}}-related energy storage}

\begin{figure}
	\includegraphics{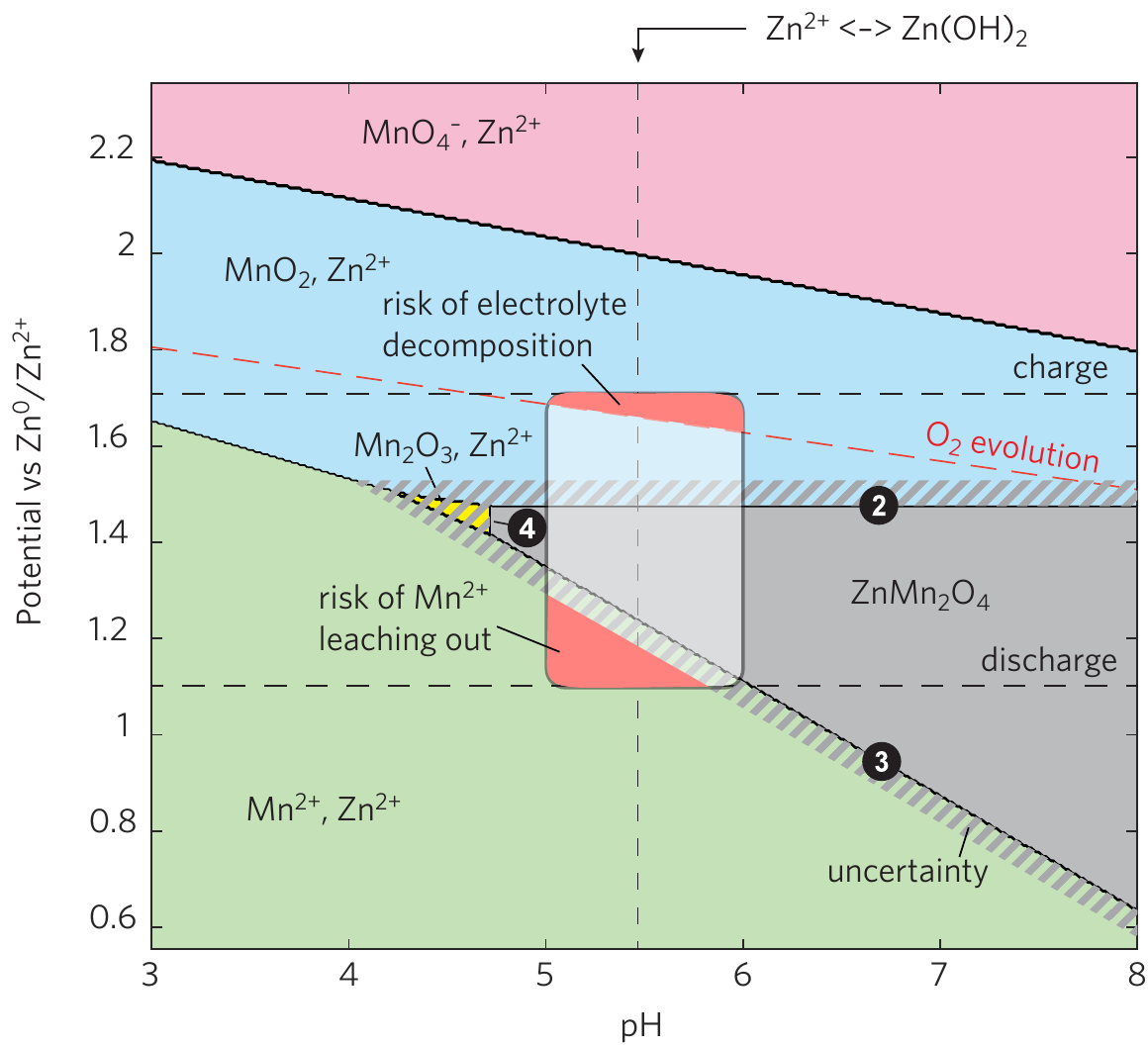}
	\caption{\ce{Mn-Zn-H2O} Pourbaix diagram with \ce{ZnMn2O4} phase boundaries. The battery operating window is shown as a wide bar. The hatched area reflects the uncertainty in the stability region of \ce{ZnMn2O4} due to the uncertainty in its chemical potential (Table~\ref{tab-Chem-pot}). Phase boundaries marked with numbers are associated with the corresponding chemical reactions in the main text. The following concentrations of aqueous species are assumed $\ce{[Mn^{2+}]}=0.1$~M, $\ce{[Zn^{2+}]}=2$~M, and $\ce{[MnO4^{-}]}=0.1$~M.}\label{fig-Pourbaix-hetaerolite}
\end{figure}

We began construction of the \ce{Mn-Zn-H2O} Pourbaix diagram by mapping out a region of stability for \ce{ZnMn2O4} (hetaerolite) as the only structure of \ce{MnO2} with intercalated Zn for which thermodynamic properties are known. The charge and discharge of this structure with divalent \ce{Zn^{2+}} ions is the key fundamental process occurring during rechargeable battery operation, which is governed by the reaction 
\begin{equation}\label{eq:ZnMn2O4 charging -> Zn2+ and 2MnO2}
    \ce{Zn^{2+}(aq) + 2MnO2 + 2e- <-> ZnMn2O4}.
\end{equation}
Since there are no protons involved in this reaction, its equilibrium potential is $E_\text{SHE} = 0.72~\text{V}$ (see Eq.~(\ref{eq:ZnMn2O4 charging -> Zn2+ and 2MnO2 (full calc.)})) vs a standard hydrogen electrode (SHE), and it is independent of pH of the solution. The potential corresponds to the horizontal line~(\ref{eq:ZnMn2O4 charging -> Zn2+ and 2MnO2}) in Fig.~\ref{fig-Pourbaix-hetaerolite}, where the potential scale vs \ce{Zn^0/Zn^{2+}} is shown for convenience. The obtained equilibrium potential $E_{\ce{Zn^0/Zn^{2+}}}=E_\text{SHE} - (-0.76)~\text{V} \approx 1.5~\text{V}$ for this reaction agrees well with the experimental average \ce{Zn^{2+}} intercalation potential in \ce{MnO2} \cite{Canepa_CR_117_2017}. Here $-0.76$~V (Ref.~\citenum{Brady1982}, p.~536)  is the standard reversible potential that corresponds to the half cell reaction \ce{Zn(s) <-> Zn^{2+}(aq) + 2e- }. The calculated value of the equilibrium potential for reaction (\ref{eq:ZnMn2O4 charging -> Zn2+ and 2MnO2}) has a small uncertainty (ca. 0.05~V, see hatched region in Fig.~\ref{fig-Pourbaix-hetaerolite}), which comes from two slightly different values of $\mu^{\circ}(\ce{ZnMn2O4})$ in Table~\ref{tab-Chem-pot}. The formation of hetaerolite is experimentally confirmed in Zn/EMD cells after discharge at 1.35~V vs \ce{Zn^0/Zn^{2+}} \cite{Tran_SR_11_2021}, which supports our diagram (Fig.~\ref{fig-Pourbaix-hetaerolite}).

Here we use pyrolusite (\ce{$\beta$-MnO2}) as a reference for $\mu^{\circ}(\ce{MnO2})$ since it is the most stable natural form of \ce{MnO2}. Other \ce{MnO2} polymorphs are used in batteries, e.g., \ce{$\alpha$-MnO2} \cite{Alfaruqi_JPS_288_2015}, as well as $\varepsilon-$, R$-$, and \ce{$\gamma$-MnO2} as main constituents of EMD. The question is whether the structure of \ce{MnO2} makes a significant difference to the Pourbaix diagram. We are not aware of any experimentally measured energy differences between the $\beta$ phase and other polymorphs but any differences can be inferred from DFT calculations. Our results suggest that \ce{$\beta$-MnO2} has the lowest total energy followed by R$-$ and \ce{$\gamma$-MnO2} (both have an additional 0.01~eV/f.u. above \ce{$\beta$-MnO2}) and \ce{$\alpha$-MnO2} (additional 0.05~eV/f.u.). \citet{Kitchaev_PRB_93_2016} came to a similar conclusion (ca. 0.05, 0.04, and 0.07~eV/f.u. for R, $\gamma$, and $\alpha$ phases) with a more chemically accurate metha-GGA exchange correlation functional. If \ce{$\alpha$-MnO2} is used instead, the equilibrium potential for reaction (\ref{eq:ZnMn2O4 charging -> Zn2+ and 2MnO2}) would increase by 0.05~V, which is not critical for the purpose of our discussion and would fall within the error margins of experimental formation energies (see hatched region in Fig.~\ref{fig-Pourbaix-hetaerolite}).

Now we turn to dissolution of the cathode material. It is expressed by the following reaction:
\begin{equation}\label{eq:ZnMn2O4 -> Zn2+ and Mn2+ dissolution}
    \ce{ZnMn2O4 + 8H+(aq) + 2e- <-> Zn^{2+}(aq) + 2Mn^{2+}(aq) + 4H2O}.
\end{equation}
The equilibrium potential for reaction (\ref{eq:ZnMn2O4 -> Zn2+ and Mn2+ dissolution}) is $E_\text{SHE} = 1.78~\text{V} - \text{pH}\times0.237~\text{V}$ (see Eq.~(\ref{eq:ZnMn2O4 -> Zn2+ and Mn2+ dissolution (full calc.)})), which is a function of pH as the reaction involves both water and protons. The dissolution boundary is marked as~(\ref{eq:ZnMn2O4 -> Zn2+ and Mn2+ dissolution}) in Fig.~\ref{fig-Pourbaix-hetaerolite} with the uncertainty as per the discussion above. Once we put this boundary in the context of the pH and potential range for Zn/\ce{MnO2} battery operation (represented by a rectangle in Fig.~\ref{fig-Pourbaix-hetaerolite}), we immediately recognize that the lower boundary of \ce{ZnMn2O4} electrochemical stability is close to the lower boundary of the discharge potential. This result is in line with \citet{Li_CM_31_2019} who attributed the capacity fade to a dissolution of the active cathode material into the electrolyte that takes place at lower voltages (less than 1.26~V vs \ce{Zn^0/Zn^{2+}}). \citet{Tran_SR_11_2021} reached a similar conclusion after experimental studies involving detection of soluble \ce{Mn^{2+}}
species formed during discharge, namely that \ce{Mn^{2+}} is formed/dissolved at lower voltages (below $\sim 1.1$~V vs \ce{Zn^0/Zn^{2+}}).

We can also see from the Pourbaix diagram in Fig.~\ref{fig-Pourbaix-hetaerolite} that a more acidic electrolyte (i.e., lower pH) favours cathode dissolution ($\text{pH}<5$). At the same time, pH is buffered by the formation of \ce{Zn(OH)2} ($\text{pH}>5.5$). This leaves us with a narrow pH range of $5<\text{pH}<5.5$ and a limited lowest discharge potential of $E_{\ce{Zn^0/Zn^{2+}}}^{\text{min}}=1.2$~V to avoid cathode corrosion.

To complete the map of \ce{ZnMn2O4} electrochemical stability, we also consider its transformation into \ce{Mn2O3}
\begin{equation}\label{eq:ZnMn2O4 -> Mn2O3}
    \ce{ZnMn2O4 + 2H+ <-> Zn^{2+}(aq) + Mn2O3 + H2O}.
\end{equation}
Here the transition occurs at a fixed $\text{pH} = 4.75$ (line~(\ref{eq:ZnMn2O4 -> Mn2O3}) in Fig.~\ref{fig-Pourbaix-hetaerolite}) since the reaction does not involve electrons (see Eq.~(\ref{eq:ZnMn2O4 -> Mn2O3 (full calc.)})). This pH boundary value is extremely sensitive to chemical potentials of species involved in the reaction. The \ce{Mn2O3} region can even vanish (see hatched zone in Fig.~\ref{fig-Pourbaix-hetaerolite}) with 0.1~eV uncertainty in $\mu^{\circ}(\ce{ZnMn2O4})$. This marks the pH boundary below which \ce{ZnMn2O4} is unstable and, thus, energy storage due to \ce{Zn^{2+}} intercalation is not possible.

\subsection{Chalcophanite}

\begin{figure}
	\includegraphics{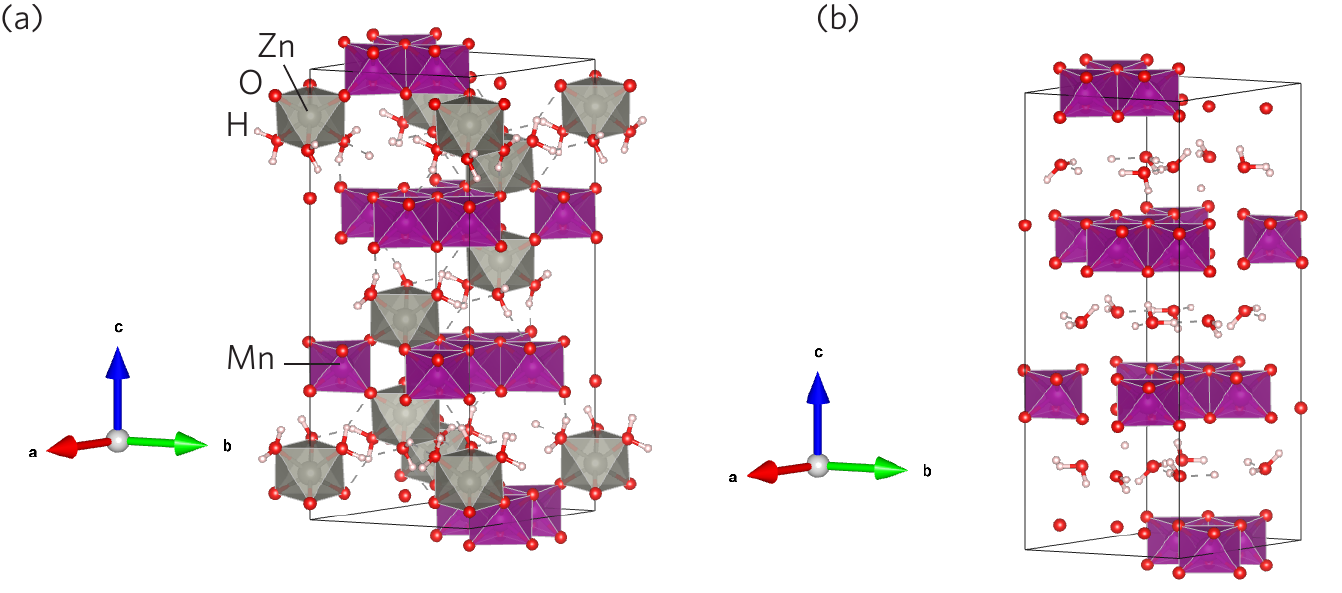}
	\caption{(a) Chalcophanite \ce{ZnMn3O7.3H2O} and (b) chalcophanite with deintercalated zinc \ce{Mn3O7.3H2O}.}\label{fig-chalcophanite-Zn-deintercalation}
\end{figure}

Chalcophanite (\ce{ZnMn3O7.3H2O}) is a byproduct phase formed in the cathode material of rechargeable aqueous \ce{Zn/MnO2} batteries upon cycling (see Sec.~Introduction). To understand the role of \ce{ZnMn3O7.3H2O} in energy storage we begin with a review of its crystal structure, as shown in Fig.~\ref{fig-chalcophanite-Zn-deintercalation}(a). The stoichiometry can be viewed as an \ce{Mn_{0.86}O2} manganese-deficient layered structure where two Zn atoms electronically compensate for each Mn vacancy. At first glance, we can expect the chalcophanite phase to participate in the energy storage similar to \ce{ZnMn2O4} where \ce{Zn^{2+}} can be deintercalated by an applied potential. The question that comes to mind is regarding the magnitude of the potential. If the thermodynamic properties of \ce{ZnMn3O7.3H2O} were experimentally determined, we would follow the same approach used in the analysis of reaction (\ref{eq:ZnMn2O4 charging -> Zn2+ and 2MnO2}). However, in the absence of experimental data we can only rely on DFT calculations.

\citet{Aydinol_PRB_56_1997} proposed a method for calculation of an average Li intercalation voltage $\bar{E}$ in transition metal oxides based on DFT total energies $E$. The original expression modified for the case of \ce{ZnMn3O7.3H2O} is
\begin{equation}\label{eq:intercalation potential DFT}
    \bar{E}_{\ce{Zn^0/Zn^{2+}}} = 
    \frac{H(\ce{Zn_{$x_1$}Mn3O7.3H2O}) + 
    (x_2 - x_1)H(\ce{Zn}) -
    H(\ce{Zn_{$x_2$}Mn3O7.3H2O})}{2e(x_2 - x_1)},
\end{equation}
where $H(\ce{Zn})$ is the DFT total energy of bulk Zn, $e$ is the magnitude of the electron charge, and $x_2$ and  $x_1$ are the effective number of Zn atoms in the structure before and after deintercalation ($x_1 < x_2$). The factor of two in the denominator reflects the ionic charge of $\ce{Zn^{2+}}$. Our structural model of \ce{ZnMn3O7.3H2O} ($x_2=1$) is based on experimental data reported by \citet{Post_AM_73_1988} (space group 148, $R\bar{3}$ hexagonal axes) with lattice parameters $a=b=7.53-7.54$ and $c=20.79-20.82$~{\AA}. DFT lattice parameters $a=b=7.54$ and $c=20.51$~{\AA} were obtained after  full structural relaxation. The structure of \ce{Mn3O7.3H2O} ($x_1=0$) with deintercalated Zn (Fig.~\ref{fig-chalcophanite-Zn-deintercalation}b) was obtained by removing Zn followed by full structural relaxation. The total energies of these two structures as well as the bulk Zn were used to calculate the intercalation potential using Eq.~(\ref{eq:intercalation potential DFT}).

The DFT-predicted Zn intercalation potential in \ce{ZnMn3O7.3H2O} is $\bar{E}_{\ce{Zn^0/Zn^{2+}}}=2.6$~V vs \ce{Zn^0/Zn^{2+}}. This is a very large value compared to 1.5~V for \ce{ZnMn2O4}. It should be noted that a DFT with PBE exchange-correlation functional always \textit{underestimates} redox potentials relative to experiments due to its failure to properly capture correlation effects on the transition metal ion \cite{Chevrier_PRB_82_2010}. Such a large Zn intercalation potential in \ce{Zn_{1/0}Mn3O7.3H2O} can be attributed to the \ce{Mn^{4+/5+}} redox reaction instead of \ce{Mn^{3+/4+}} as in the case of \ce{Zn_{1/0}Mn2O4}.

\begin{figure}
	\includegraphics{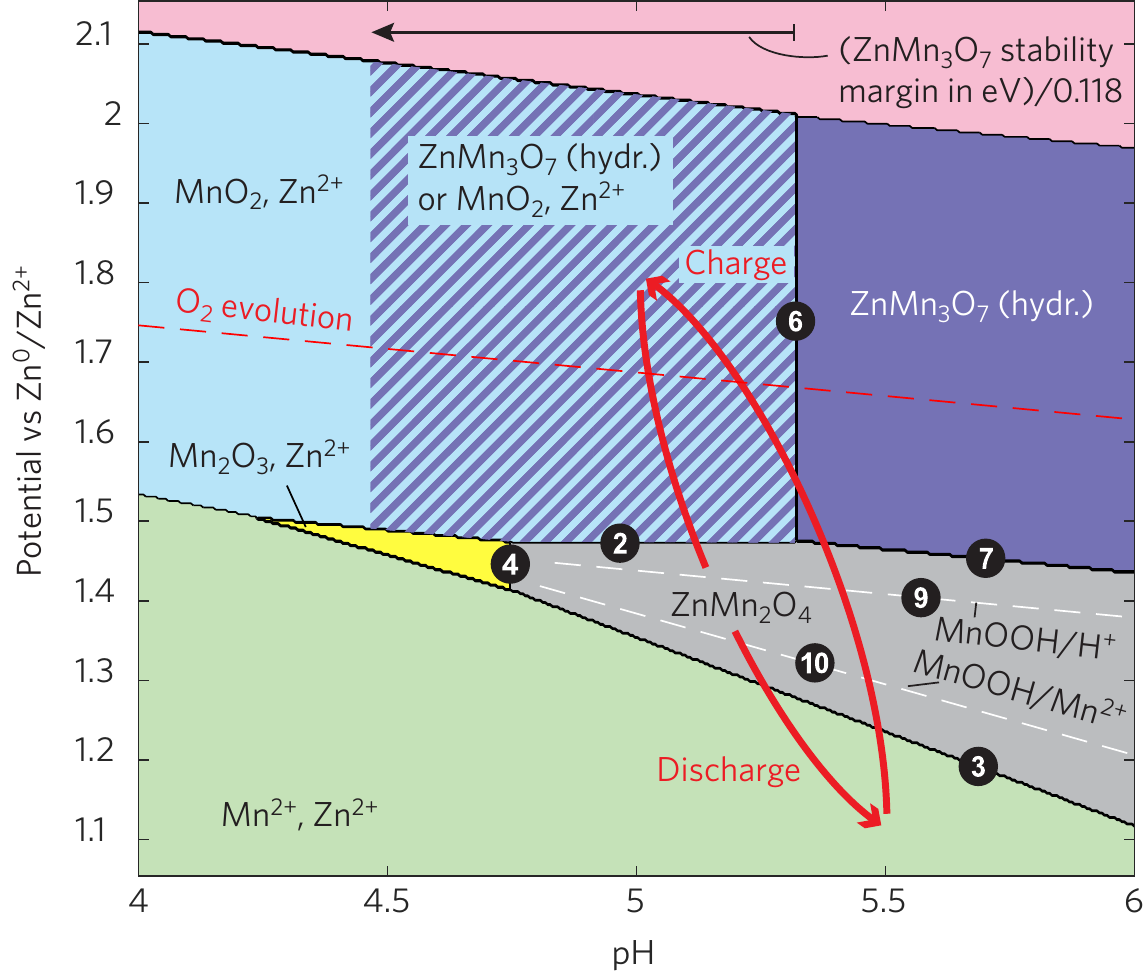}
	\caption{\ce{Mn-Zn-H2O} Pourbaix diagram with boundaries for \ce{ZnMn2O4}, \ce{ZnMn3O7}~(hydr.), and \ce{MnOOH} phases. The hatched region represent an uncertainty in \ce{ZnMn3O7}~(hydr.) stability margin $\delta\mu=0-0.1$~eV/f.u. A typical charge-discharge cycle of a \ce{Zn/MnO2} cell is shown schematically. Phase boundaries marked with numbers are associated with the corresponding chemical reactions in the main text. The \ce{Zn(OH)2} stability boundary at pH~5.45 is not shown. The following concentrations of aqueous species are assumed $\ce{[Mn^{2+}]}=0.1$~M, $\ce{[Zn^{2+}]}=2$~M, and $\ce{[MnO4^{-}]}=0.1$~M. }\label{fig-Pourbaix-hetaerolite-chalcophinite}
\end{figure}

To add \ce{ZnMn3O7.3H2O} to the Pourbaix diagram we need to identify its boundaries of stability. This task is complicated by the lack of its experimental chemical potential, and DFT calculations are still too inaccurate for this purpose. Thus we assume $\mu^{\circ}(\ce{ZnMn3O7}~\text{hydr.}) = \mu^{\circ}(\ce{ZnO}~\text{hydr.}) + 3\mu^{\circ}(\ce{MnO2}) - \delta\mu$, where $\delta\mu$ is a stability margin which will be kept as a variable. First we investigate \ce{ZnMn3O7} decomposition with \ce{MnO2} as one of the products
\begin{equation}\label{eq:ZnMn3O7 -> MnO2 + Zn2+}
    \ce{ZnMn3O7 + 2H+(aq) <-> Zn^{2+}(aq) + 3MnO2 + H2O}.
\end{equation}
This reaction has no charge transfer and, thus, cannot be used for energy storage. The equilibrium boundary is at $\text{pH} = 5.32 - 0.118^{-1}\delta\mu$ ($\delta\mu$ should be in eV). Figure~\ref{fig-Pourbaix-hetaerolite-chalcophinite} presents the Pourbaix diagram with chalcophanite. The equilibrium boundary for this reaction, line~(\ref{eq:ZnMn3O7 -> MnO2 + Zn2+}), is included with the uncertainty region that reflects $\delta\mu = 0-0.1$~eV. We believe that this is a correct order of magnitude for $\delta\mu$ as we did not observe chalcophanite experimentally at low pH values (see Sec.~Experimental validation).

To map a boundary between chalcophanite and \ce{ZnMn2O4} we investigated the following reaction
\begin{equation}\label{eq:ZnMn3O7 -> ZnMn2O4 + Zn2+}
    \ce{3ZnMn2O4 + 2H2O <-> 2ZnMn3O7 + Zn^{2+}(aq) + 4H+ + 6e^-}
\end{equation}
that yields $E_\text{SHE} = 0.93 - \delta\mu/3\text{e} - 0.0395\,\text{pH}$ (line~(\ref{eq:ZnMn3O7 -> ZnMn2O4 + Zn2+}) in Fig.~\ref{fig-Pourbaix-hetaerolite-chalcophinite}). Here $\delta\mu$ has a marginal influence on the result. At pH~5.5 the boundary will be at 0.71~V vs SHE or 1.47~V vs \ce{Zn^0/Zn^{2+}}, which barely affects the \ce{ZnMn2O4} stability range.

It is now possible to rationalize experimental observations related to chalcophanite in \ce{Zn/MnO2} cells and its relation to capacity fading. The region of \ce{ZnMn3O7.3H2O} electrochemical stability on the Pourbaix diagram is in the location that was roughly outlined by \citet{Huang_NL_11_2019} for \ce{Zn2Mn3O8}. \citet{Tran_SR_11_2021} also observed formation of chalcophanite after charging of Zn/EMD cells and maintaining the potential at 1.8~V vs \ce{Zn^0/Zn^{2+}} (hatched region in Fig.~\ref{fig-Pourbaix-hetaerolite-chalcophinite}) for 2~h. Most likely, reaction~(\ref{eq:ZnMn3O7 -> MnO2 + Zn2+}) is kinetically too slow and, instead, chalcophanite is deposited from dissolved aqueous species during battery charging~\cite{Tran_SR_11_2021} (hatched region in Fig.~\ref{fig-Pourbaix-hetaerolite-chalcophinite})
\begin{equation}\label{eq:MnO2 + Zn2+ -> ZnMn3O7}
    \ce{Zn^{2+}(aq) + 3Mn^{2+}(aq) + 7H2O -> ZnMn3O7 + 14H+ + 6e-}.
\end{equation}
Here the presence of \ce{Mn^{2+}(aq)} in the electrolyte comes from the cathode material previously leached out during discharge, which explains the connection between chalcophanite formation and a deep discharge potential established by \citet{Li_CM_31_2019}. Thus, chalcophanite acts as a scavenger in consuming dissolved \ce{Mn^{2+}(aq)} species preventing them from depositing back as \ce{MnO2}.

Here we have not conducted calculations on the \ce{Zn2Mn3O8} compound since we do not have enough knowledge about its structure, thermodynamics, and experimental evidence for its importance in battery cycling performance. Also, we have not discussed zinc hydroxide sulfate \ce{Zn4(OH)6(SO4).$x$H2O}; the relevant Pourbaix diagram of \ce{Zn-S-H2O} has been reported elsewhere \cite{Tran_SR_11_2021}.

\subsection{Complementary energy storage mechanisms}

It is widely discussed in the literature \cite{Wen_EA_50_2004,Oberholzer_AAMI_11_2018,Zhao_S_15_2019,Li_CM_31_2019} that the proton exchange reaction
\begin{equation}\label{eq:MnOOH -> MnO2 + H+}
    \ce{MnO2 + H+ + e- <-> MnOOH}
\end{equation}
has a significant contribution to energy storage in \ce{Zn/MnO2} aqueous cells. Interestingly, \ce{MnOOH} was not included in the original \ce{Mn-H2O} Pourbaix diagram \cite{Pourbaix1974}, and its chemical potential was omitted.  \citet{Bratsch_JPCRD_18_1989} reported that $E^{\circ}=0.98$~V for this reaction with \ce{$\beta$-MnO2} at $\text{pH}=0$, which translates into $\mu^{\circ}(\ce{MnOOH})=-5.81$~eV/f.u. (There is an alternative value of $\mu^{\circ}(\ce{MnOOH})$ listed in Table~\ref{tab-Chem-pot} that indicates the amount of scatter in the experimental data.)

According to Eq.~(\ref{eq:MnOOH -> MnO2 + H+ (full calc.)}) the corresponding \ce{H+} deintercalation potential in  reaction~(\ref{eq:MnOOH -> MnO2 + H+}) is $E_\text{SHE} = 0.98 - 0.0592\,\text{pH}$ (in V), which amounts to $E_{\ce{Zn^0/Zn^{2+}}}=1.41$~V at pH~5.5. This potential is very close to $E_{\ce{Zn^0/Zn^{2+}}} \approx 1.5~\text{V}$ for the \ce{Zn^{2+}} deintercalation reaction, as shown in Fig.~\ref{fig-Pourbaix-hetaerolite-chalcophinite} by lines (\ref{eq:ZnMn2O4 charging -> Zn2+ and 2MnO2}) and (\ref{eq:MnOOH -> MnO2 + H+}). \citet{Magar_JES_167_2020} arrived at the same result via DFT calculations, namely suggesting that proton insertion is only 0.1~eV less favourable than \ce{Zn^{2+}} intercalation per \ce{MnO2} formula unit. Since both energy mechanisms (\ce{Zn^{2+}} and \ce{H^{+}} deintercalation) require nearly the same voltage, we can conclude that both can coexist and contribute to energy storage in the \ce{Zn/MnO2} aqueous cell.  Cyclic voltammetry \cite{Zhao_S_15_2019} of \ce{Zn/MnO2} cells also revealed two redox peaks upon both charge and discharge; the peaks were spaced $0.1-0.2$~V apart (the same as our calculated equilibrium potentials for reactions (\ref{eq:ZnMn2O4 charging -> Zn2+ and 2MnO2}) and (\ref{eq:MnOOH -> MnO2 + H+}) in Fig.~\ref{fig-Pourbaix-hetaerolite-chalcophinite}) and are attributed to two redox reactions that involve \ce{Zn^{2+}/H^+} intercalation in \ce{MnO2}. It should be noted that the potential corresponding to reduction and oxidation peaks is sensitive ($\pm 0.1$~V) to the crystal structure (polymorph) of \ce{MnO2}, presence of defects, particle size and morphology \cite{Guo_N_8_2018,Chamoun_ESM_15_2018,Huang_NC_9_2018,Khamsanga_SR_9_2019,Han_iS_23_2020,Chen_MCF_4_2020,Zhou_FER_8_2020,Zeng_V_192_2021}. This implies that conclusions drawn from the Pourbaix diagram are robust with respect to those extrinsic factors.

\ce{MnOOH} can also be further reduced to \ce{Mn^{2+}(aq)}
\begin{equation}\label{eq:MnOOH -> Mn2+}
    \ce{MnOOH + 3H+ + e- <-> Mn^{2+}(aq) + 2H2O},
\end{equation}
which takes place at $E_\text{SHE} = 1.52 - 0.178\,\text{pH}$ (in V) as shown in Fig.~\ref{fig-Pourbaix-hetaerolite-chalcophinite}, line~(\ref{eq:MnOOH -> Mn2+}). At pH~5.5 this amounts to an equilibrium potential of $E_{\ce{Zn^0/Zn^{2+}}}=1.31$~V  slightly above that for the \ce{ZnMn2O4} dissolution/reduction reaction (see line~(\ref{eq:ZnMn2O4 -> Zn2+ and Mn2+ dissolution}) in Fig.~\ref{fig-Pourbaix-hetaerolite-chalcophinite}). This suggests that \ce{MnOOH} is more susceptible to dissolution/reduction than \ce{ZnMn2O4} during discharge. Interestingly, 1.31~V is remarkably close to 1.26~V vs \ce{Zn^0/Zn^{2+}} which is marked as a boundary for capacity fade reactions by \citet{Li_CM_31_2019}.

The cathode dissolution process (if reversible) can also participate in the energy storage via the following reaction
\begin{equation}\label{eq:MnO2 -> Mn2+}
    \ce{MnO2 + 4H+ + 2e- <-> Mn^{2+}(aq) + 2H2O}.
\end{equation}
This reaction involves \ce{2e-}, while the conventional Zn intercalation reaction~\eqref{eq:ZnMn2O4 charging -> Zn2+ and 2MnO2} or the proton intercalation reaction~\eqref{eq:MnOOH -> MnO2 + H+} involves only \ce{1e-} per \ce{MnO2} formula unit. Thus, the \ce{MnO2} cathode dissolution results in a theoretical specific capacity of nearly 600~mA~h~g$^{-1}$. This is, however, quite an unrealistic figure assuming that the entire cathode dissolves in the electrolyte. \citet{Lee_SR_4_2014} presented experimental evidence for reversible dissolution of 1/3 of the \ce{$\alpha$-MnO2} cathode material during the charge-discharge cycle. According to this scenario, half of the energy storage capacity would come from Zn intercalation, and the other half would be contributed by cathode dissolution. \citet{Moon_AS_8_2021} support this argument by finding \textit{ex situ} a substantial amount of \ce{$\alpha$-MnO2}
remaining unreacted even after full discharge even though the cell delivered ca. 90\% of its theoretical specific capacity for \ce{Zn^{2+}} or \ce{H+} intercalation. References~\citenum{Tran_SR_11_2021} and \citenum{Moon_AS_8_2021} also provide experimental evidence for electrochemical co-deposition of chalcophanite from \ce{Mn^{2+}(aq)} and \ce{Zn^{2+}(aq)} species at the full charge state and its subsequent dissolution during discharge, which agrees with the Pourbaix diagram in Fig.~\ref{fig-Pourbaix-hetaerolite-chalcophinite}.

\subsection{Experimental validation}\label{Sec:Results:Experimental validation}

To validate the \ce{Zn-Mn-H2O} Pourbaix diagram in Fig.~\ref{fig-Pourbaix-hetaerolite-chalcophinite} we constructed a set of experiments that involve \textit{ex situ} characterization of the electrolytic \ce{MnO2} cathode material after cycling. Different regions of the Pourbaix diagram were sampled by changing the acidity of the electrolyte (pH~2.5 and 4). These pH values reflect the `as prepared' electrolyte. During battery operation, the pH of the electrolyte evolves as a result of an increase in \ce{OH-} concentration during discharge followed by zinc hydroxide sulphate (ZHS) precipitation that buffers the pH to $\sim5.5$~\cite{Lee_C_9_2016,Chamoun_ESM_15_2018,Bischoff_JES_167_2020}. When the initial pH of the electrolyte was 4, ZHS started to appear at potentials below 1.35~V but before 1.18~V. For the electrolyte with an initial pH of 2.5, ZHS formation was suppressed and did not occur until after a potential of 1.18~V was reached.

\begin{figure}
	\includegraphics{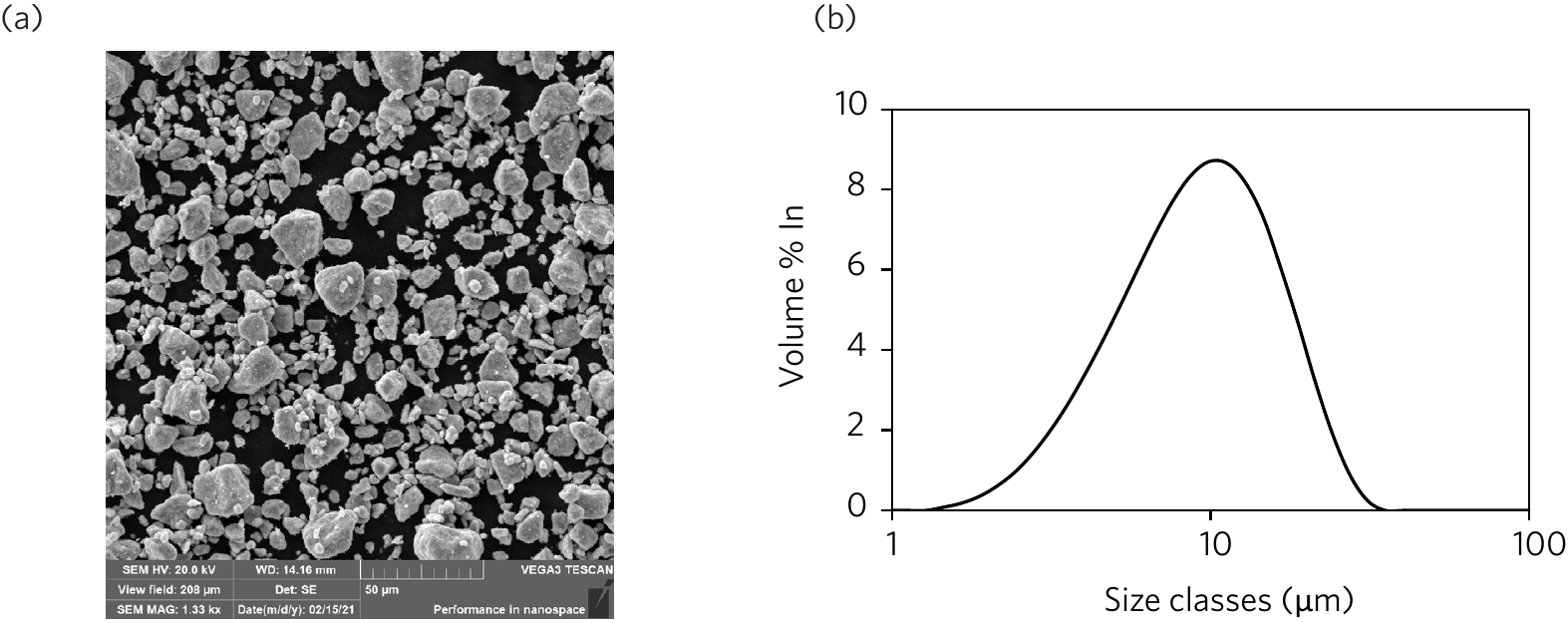}
	\caption{EMD particles: (a) SEM secondary electron image, (b) particle size distribution.}\label{fig-MnO2-particle-size}
\end{figure}

\begin{figure}
	\includegraphics{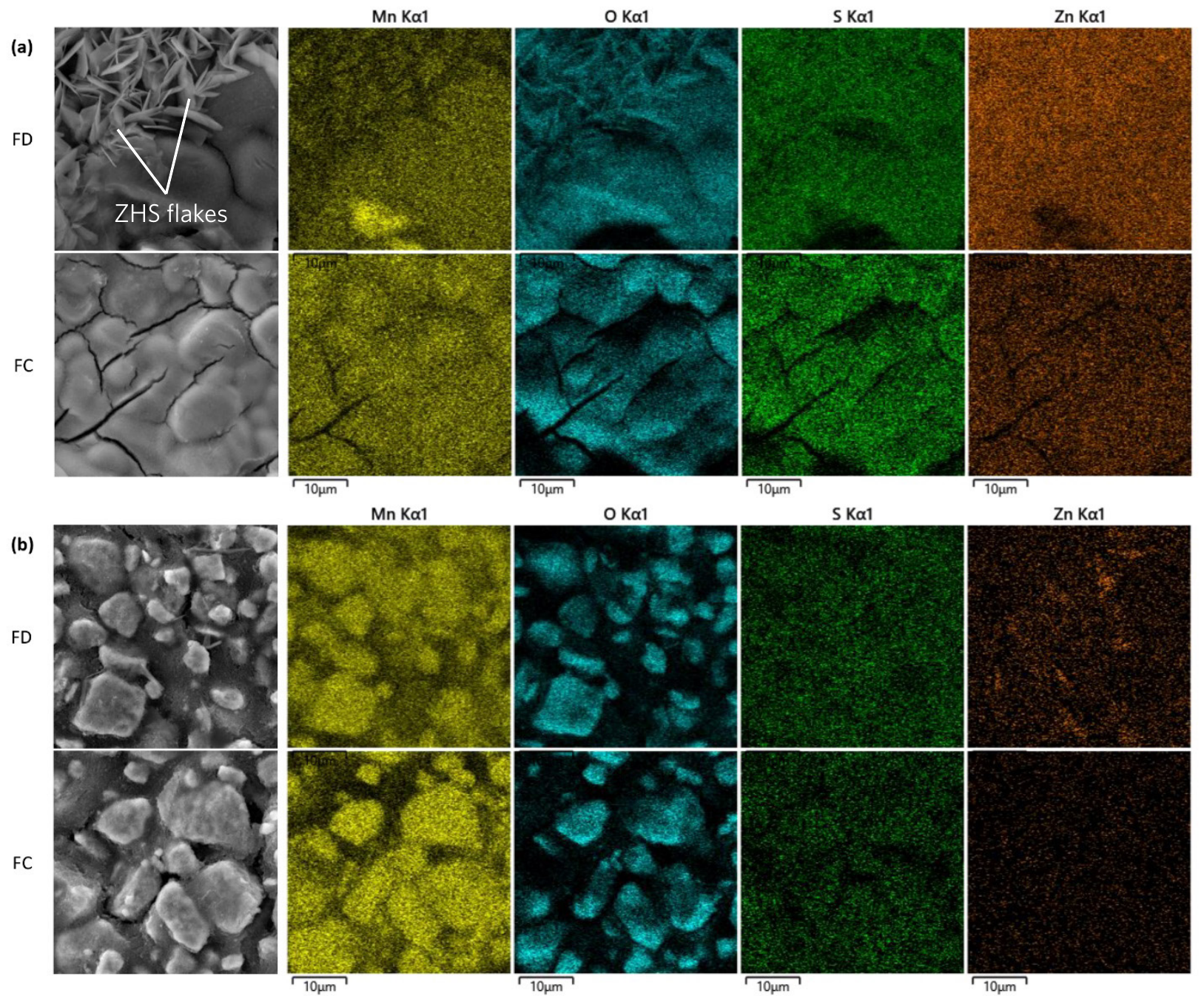}
	\caption{SEM SE images and corresponding EDX maps of EMD electrodes after 5 cycles in electrolytes with different pH values: (a) pH 4 and (b) pH 2.5. Fully-discharged and fully-charged states are denoted as FD and FC, respectively.}\label{fig-SEM-EDX}
\end{figure}

Figure~\ref{fig-MnO2-particle-size} shows EMD particles and their size distribution prior to electrode fabrication. There is a fairly wide distribution of particle size with a mean size of ca.~10~$\mu$m. Morphological changes of EMD electrodes after 5 cycles in electrolytes with different pH values are shown in Fig.~\ref{fig-SEM-EDX}. For pH~4 (Fig.~\ref{fig-SEM-EDX}a), during discharge, the electrode surface is covered with ZHS flakes and a new layer of Zn-Mn oxide, consisting of hetaerolite and Zn-intercalated Mn oxide. During charge, ZHS disappears while the new Zn-Mn-oxide layer remains. This new layer has a characteristic `desert crack pattern' without distinct particles; its morphology is very different from the original EMD electrode (see Ref.~\citenum{Tran_SR_11_2021}, Fig.~6a therein). This suggests that this Zn-Mn-oxide layer was electrochemically deposited during cycling. More detailed SEM characterization of the EMD surface morphology before and after cycling including various charge and discharge depths is presented in Ref.~\citenum{Tran_SR_11_2021}.

For pH~2.5 (Fig.~\ref{fig-SEM-EDX}b), the morphology of the EMD electrode surface is retained with little or no change. During discharge, a small amount of ZHS forms and its formation is reversible during charge. In general, all EDX maps show that S signal overlaps with Zn; however, Zn has covered more of the electrode surface after cycling for the pH~4 solution than the pH~2.5 solution.

\begin{figure}
	\includegraphics{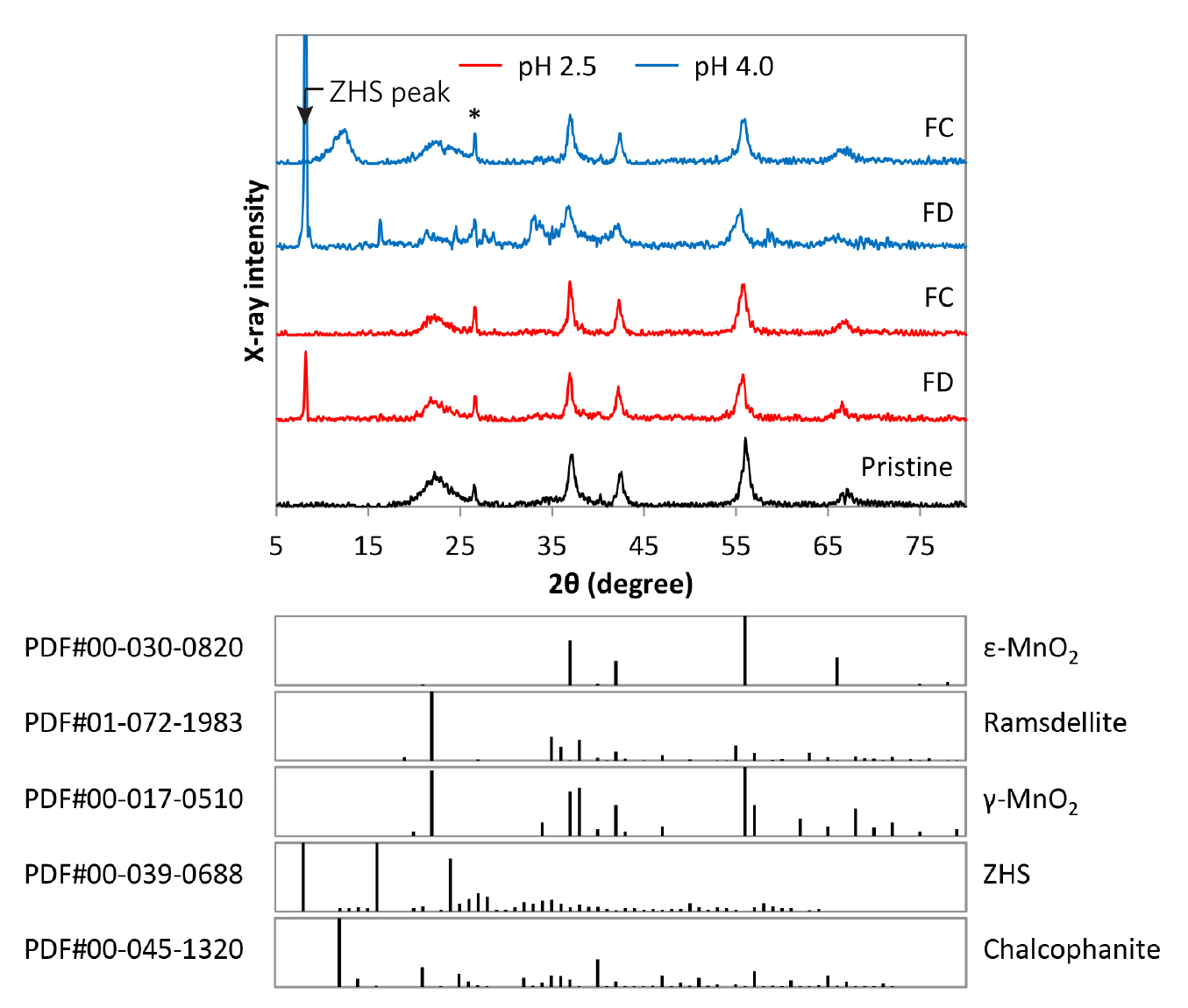}
	\caption{XRD patterns of EMD electrodes after 5 cycles during GCD tests in electrolytes with different pH values. Reference PDF cards for $\varepsilon$-MnO2, ramsdellite, $\gamma$-MnO2, ZHS, and chalcophanite are included at the bottom of the XRD patterns. Fully-discharged and fully-charged states are denoted as FD and FC, respectively. The asterisk indicates a graphite peak from the current collector.}\label{fig-XRD}
\end{figure}

Pristine EMD has been characterized in Ref.~\citenum{Tran_SR_11_2021}. Briefly, the EMD powder is composed of about 53\% \ce{$\varepsilon$-MnO2}, 34\% ramsdellite, and 13\% \ce{$\gamma$-MnO2}, determined through Rietveld refinement. XRD results of EMD electrodes after 5 cycles are shown in Fig.~\ref{fig-XRD}. For both discharged electrodes (pH 4.0 and 2.5), a new peak appears at $\sim8.1^{\circ}$ and corresponds to the formation of ZHS. The ZHS peak is significantly more intense for the pH 4.0 sample compared with the pH 2.5 sample, which indicates that more ZHS forms at the higher pH and corroborates the SEM results. During charge, the ZHS peaks disappear.

\begin{figure}
	\includegraphics{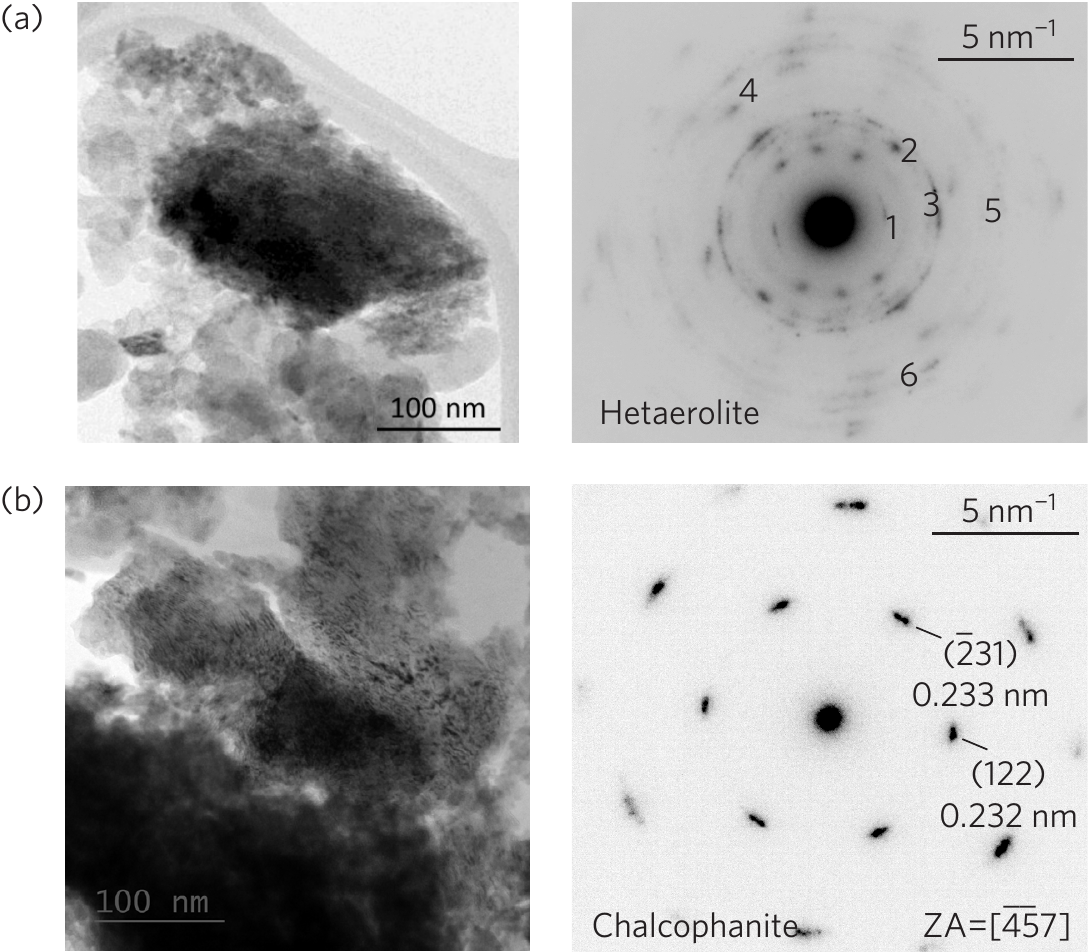}
	\caption{TEM bright field images and SAED patterns from the EMD electrodes at different stages during cycling at pH 4. (a) Electrode discharged at 1.35 V during the 1st cycle and (b) fully charged electrode held for 2 h at 1.8 V during the 50th cycle. The labels $1\!-\!6$ on the SAED pattern for hetaerolite correspond to the d-spacings, in order, in Table~\ref{tab-d-spacings}.}\label{fig-TEM-SAED}
\end{figure}

During discharge of the electrodes, the major peak ((102) plane) for \ce{$\varepsilon$-MnO2} at 56$^{\circ}$ shifts to a lower angle of 55.8$^{\circ}$ and 55.5$^{\circ}$ at pH~2.5 and 4, respectively. All other major \ce{$\varepsilon$-MnO2} peaks ((100), (101) and (110), originally at 37.1$^{\circ}$, 42.5$^{\circ}$ and 67.0$^{\circ}$, respectively) also shift to lower angles during discharge. The (102) peak shifts back to 55.9$^{\circ}$ after charging, meaning that the structure of \ce{$\varepsilon$-MnO2} can expand and collapse almost reversibly to accommodate \ce{Zn^{2+}} ions. In addition, for the charged electrode cycled in the electrolyte with a pH of~4, a new peak appears at 12.7$^{\circ}$ which matches the major peak for chalcophanite. This correlates well with the SEM images and EDX mapping of the Mn-Zn oxide layer that builds up on the EMD electrode surface during cycling in the electrolyte with pH~4. This result also matches the Pourbaix diagram (Fig.~\ref{fig-Pourbaix-hetaerolite-chalcophinite}) with the pH buffering to 5.5 and a charge potential of 1.8~V vs \ce{Zn^0/Zn^{2+}}. Chalcophanite is not observed for the electrolyte with pH~2.5, which is also expected from the Pourbaix diagram. The XRD results were not able to confirm the formation of hetaerolite (main peak is (211) at $\sim 36.5^{\circ}$) during discharge likely due to significant interference from both \ce{MnO2} and ZHS peaks. However, the presence of hetaerolite was confirmed by electron diffraction in the TEM (Fig.~\ref{fig-TEM-SAED}(a)).

According to \citet{Chamoun_ESM_15_2018}, hetaerolite should not form if the EMD electrode is cycled in a more acidic electrolyte. Thus, only electrodes discharged at 1.35~V during the 1st cycle and fully charged at 1.8~V for 2~h during the 50th cycle in pH~4 are shown in Fig.~\ref{fig-TEM-SAED}, including examples of SAED patterns from the particles shown. It should be noted that when \ce{Zn/MnO2} batteries discharge, the pH of the electrolyte in the vicinity of the EMD electrodes should increase significantly. The SAED patterns in Fig.~\ref{fig-TEM-SAED}a and b were indexed to hetaerolite (\ce{ZnMn2O4}) (Table~\ref{tab-d-spacings}) and chalcophanite (\ce{ZnMn3O7}), respectively. These results agree with the Pourbaix diagram  in Fig.~\ref{fig-Pourbaix-hetaerolite-chalcophinite}.

\begin{table}[t]
\caption{Indexed d-spacings for the SAED pattern in Figure~\ref{fig-TEM-SAED}a and d-spacings from the PDF card for hetaerolite (PDF\#01-077-0407).}
\label{tab-d-spacings}
\begin{tabular}{ l c c }
	\hline
    d-spacing ({\AA}) & Hetaerolite & $(hkl)$ \\
	\hline
    4.84 & 4.86 & (101) \\
    2.72 & 2.71 & (103) \\
    2.47 & 2.46 & (211) \\
    1.81 & 1.80 & (204) \\
    1.57 & 1.56 & (321) \\
    1.54 & 1.52 & (224) \\
    \hline
\end{tabular}
\end{table}

\section{Conclusions}

We have presented a refined \ce{Mn-Zn-H2O} Pourbaix diagram with emphasis on the battery-relevant range of parameters (pH~$4-6$, $E_{\ce{Zn^0/Zn^{2+}}}=1.1-1.8$~V). The diagram maps out boundaries of electrochemical stability for \ce{MnO2}, \ce{ZnMn2O4}, \ce{ZnMn3O7}, and \ce{MnOOH}. The diagram helps to rationalize experimental observation of processes and phases occurring during the charge/discharge of \ce{Zn/MnO2} aqueous cells. The average charge potential of 1.5~V vs \ce{Zn^0/Zn^{2+}} and the mid potential value of the redox couple peaks ($\sim 1.4$~V vs \ce{Zn^0/Zn^{2+}}) measured during cyclic voltammetry agree with the boundaries between \ce{ZnMn2O4} and \ce{MnO2} for \ce{Zn^{2+}}  intercalation ($E_{\ce{Zn^0/Zn^{2+}}}=1.5$~V) or the boundary between \ce{MnOOH} and \ce{MnO2} for \ce{H+} intercalation  ($E_{\ce{Zn^0/Zn^{2+}}}=1.4$~V at pH~5.5). The experimentally observed capacity fade attributed to dissolution of \ce{ZnMn2O4}, \ce{MnOOH}, or \ce{MnO2} at low discharge potentials (below $1.1-1.2$~V vs \ce{Zn^0/Zn^{2+}}) agrees with the boundaries between these phases and aqueous \ce{Mn^{2+}} ($E_{\ce{Zn^0/Zn^{2+}}}\approx 1.2$~V at pH~5.5). At the same time, dissolution of the cathode material can make a sizable contribution to the specific capacity of the battery (up to 50\%). The precipitation of chalcophanite \ce{ZnMn3O7} in a cathode material charged at 1.8~V correlates with its predicted range of stability above $E_{\ce{Zn^0/Zn^{2+}}}\approx 1.5$~V at pH~5.5. Chalcophanite (formed at high voltages during charge) participates in capacity fade by consuming dissolved \ce{Mn^{2+}} species and preventing their return back to \ce{MnO2}. Unlike \ce{ZnMn2O4}, it is not feasible to deintercalate \ce{Zn^{2+}} from \ce{ZnMn3O7}.

The impact of electrolyte pH has been investigated experimentally for aqueous zinc-ion \ce{MnO2} batteries. For an electrolyte with pH 4, a fraction of pristine electrolytic manganese dioxide transforms to hetaerolite and chalcophanite during discharge and charge, respectively. The transformation does not occur when the pH of electrolyte is adjusted to 2.5, which validates the proposed \ce{Mn-Zn-H2O} Pourbaix diagram.

Making progress in understanding the reaction mechanisms responsible for capacity fade can provide insight to future researchers working on performance improvements to \ce{MnO2} type cathodes. In addition, the electrochemical stability should be used as an additional design criterion in exploration of future cathode materials for aqueous rechargeable batteries.

\section{Appendix: Thermodynamic calculations of equilibrium potentials}

Here we show a detailed workflow to illustrate calculations of the equilibrium potential for the electrochemical reaction using thermodynamic data from Table~\ref{tab-Chem-pot} and assuming concentrations of aqueous species at 2~M for \ce{Zn^{2+}} and 0.1~M for \ce{Mn^{2+}}. Results are shown for the reaction (\ref{eq:ZnMn2O4 charging -> Zn2+ and 2MnO2})
\begin{equation}\label{eq:ZnMn2O4 charging -> Zn2+ and 2MnO2 (full calc.)}
\begin{gathered}
    \mu(\ce{Zn^{2+}},\text{aq}) + 2\mu^{\circ}(\ce{MnO2}) + 2\mu(\ce{e-}) = \mu^{\circ}(\ce{ZnMn2O4})\\
    -11.17~\text{eV} - 2\text{e}\,E_\text{SHE} = -12.61~\text{eV}\\
    E_\text{SHE} = 0.72~\text{V},
\end{gathered}
\end{equation}
reaction (\ref{eq:ZnMn2O4 -> Zn2+ and Mn2+ dissolution})

\begin{equation}\label{eq:ZnMn2O4 -> Zn2+ and Mn2+ dissolution (full calc.)}
\begin{gathered}
    \mu(\ce{Zn^{2+}},\text{aq}) + 2\mu(\ce{Mn^{2+}},\text{aq}) + 4\mu^{\circ}(\ce{H2O}) = \mu^{\circ}(\ce{ZnMn2O4}) + 8\mu(\ce{H+},\text{aq}) + 2\mu(\ce{e-}) \\
    -16.18~\text{eV} = -12.61~\text{eV} - \text{pH}\times0.476~\text{eV} - 2\text{e}\,E_\text{SHE} \\
    E_\text{SHE} = 1.78~\text{V} - \text{pH}\times0.237~\text{V},
\end{gathered}
\end{equation}
reaction (\ref{eq:ZnMn2O4 -> Mn2O3})
\begin{equation}\label{eq:ZnMn2O4 -> Mn2O3 (full calc.)}
\begin{gathered}
    \mu(\ce{Zn^{2+}},\text{aq}) + \mu^{\circ}(\ce{Mn2O3}) + \mu^{\circ}(\ce{H2O}) = \mu^{\circ}(\ce{ZnMn2O4}) + 2\mu(\ce{H+})\\
    -13.17~\text{eV} = -12.61~\text{eV} - 0.118~\text{eV} \times \text{pH}\\
    \text{pH} = 4.75,
\end{gathered}
\end{equation}
reaction (\ref{eq:ZnMn3O7 -> MnO2 + Zn2+})
\begin{equation}\label{eq:ZnMn3O7 -> MnO2 + Zn2+ (full calc.)}
\begin{gathered}
    \mu^{\circ}(\ce{ZnMn3O7}~\text{hydr.}) + 2\mu(\ce{H+}) = \mu(\ce{Zn^{2+}},\text{aq}) + 3\mu^{\circ}(\ce{MnO2}) + \mu^{\circ}(\ce{H2O})\\
    -17.83~\text{eV} - \delta\mu - 0.118~\text{eV} \times \text{pH} = -18.46~\text{eV}\\
    \text{pH} = 5.32 - \frac{\delta\mu~\text{(eV)}}{0.118},
\end{gathered}
\end{equation}
reaction (\ref{eq:ZnMn3O7 -> ZnMn2O4 + Zn2+})
\begin{equation}\label{eq:ZnMn3O7 -> ZnMn2O4 + Zn2+ (full calc.)}
\begin{gathered}
    3\mu^{\circ}(\ce{ZnMn2O4}) +  2\mu^{\circ}(\ce{H2O}) = 2\mu^{\circ}(\ce{ZnMn3O7}) + \mu(\ce{Zn^{2+}},\text{aq}) + 4\mu(\ce{H^+}) + 6\mu(\ce{e-})\\
    -42.75~\text{eV} = -37.16~\text{eV} - 2\delta\mu - 0.237~\text{eV} \times \text{pH} - 6\text{e}\,E_\text{SHE}\\
    E_\text{SHE} = 0.93 - \delta\mu/3\text{e} - \text{pH}\times0.0395~\text{V},
\end{gathered}
\end{equation}
reaction (\ref{eq:MnOOH -> MnO2 + H+})
\begin{equation}\label{eq:MnOOH -> MnO2 + H+ (full calc.)}
\begin{gathered}
    \mu^{\circ}(\ce{MnO2}) + \mu(\ce{H+},\text{aq}) + \mu(\ce{e-}) = \mu^{\circ}(\ce{MnOOH})\\
    -4.83~\text{eV} - \text{pH}\times0.0592~\text{eV} - \text{e}\,E_\text{SHE} = -5.81~\text{eV}\\
    E_\text{SHE} = 0.98 - \text{pH}\times0.0592~\text{V},
\end{gathered}
\end{equation}
reaction (\ref{eq:MnOOH -> Mn2+})
\begin{equation} \label{eq:MnOOH + 3H+ -> Mn2+ + H2O}
    \begin{gathered}
        \mu^{\circ}(\ce{MnOOH}) + 3\mu(\ce{H^+}) + \mu(\ce{e^-}) = \mu(\ce{Mn^{2+}},\text{aq}) + 2\mu^{\circ}(\ce{H2O})\\
        -5.81~\text{eV} - \text{pH}\times0.178~\text{eV} - \text{e} E_{\text{SHE}} = -7.33~\text{eV}\\
        E_{\text{SHE}} = 1.52 - \text{pH}\times0.178~\text{V}.
    \end{gathered}
\end{equation}

%
%
%

\begin{acknowledgement}

This work was supported by the Salient Energy, the NSERC Alliance program, the Mitacs Globalink program, the International Manganese Institute (IMnI) and a grant through the Mitacs Accelerate program (FR49801).

\end{acknowledgement}

\bibliography{bibliography}

\end{document}